\documentclass[12pt]{article}
\usepackage{graphicx}
\usepackage{hyperref}

\def \bA{\overline{A}}
\def \bea{\begin{eqnarray}}
\def \beq{\begin{equation}}

\def \eea{\end{eqnarray}}
\def \eeq{\end{equation}}
\def \gef{\gamma_{\rm eff}}
\def \ko{K^0}
\def \({\left(}
\def \){\right)}
\def \[{\left[}
\def \]{\right]}
\def \nn{\nonumber}

\def \od{\overline{D}^0}
\def \ok{\overline{K}^0}

\def \i{{\it i}}

\begin{document}

\rightline{EFI 12-36}
\rightline{UdeM-GPP-TH-13-219}
\rightline{TECHNION-PH-13-1}
\rightline{January 2013}

\bigskip
\centerline{\bf SHIFT IN WEAK PHASE $\gamma$ DUE TO CP ASYMMETRIES}
\centerline{\bf IN $D$ DECAYS TO TWO PSEUDOSCALAR MESONS}
\bigskip
\centerline{Bhubanjyoti Bhattacharya and David London}
\centerline{\it Physique des Particules, Universit\'e de Montr\'eal}
\centerline{\it C.P. 6128, succ.\ centre-ville, Montr\'eal, QC, Canada H3C 3J7}
\medskip

\centerline{Michael Gronau}
\centerline{\it Physics Department, Technion -- Israel Institute of Technology}
\centerline{\it Haifa 3200, Israel}
\medskip

\centerline{Jonathan L. Rosner}
\centerline{\it Enrico Fermi Institute and Department of Physics}
\centerline{\it University of Chicago, 5620 S. Ellis Avenue, Chicago, IL 60637}
\bigskip

\begin{quote}
A difference of several tenths of a percent has been observed
between the direct CP asymmetries of $D^0\to K^+K^-$ and $D^0\to \pi^+\pi^-$.
It has been noted recently that CP asymmetries in such
singly-Cabibbo-suppressed (SCS) decays can affect the determination of
the weak phase $\gamma$ using the Gronau-London-Wyler method of
comparing rates for $B^+ \to D K^+$ and $B^- \to D K^-$, where $D$ is
a superposition of $D^0$ and $\od$ decaying to a CP eigenstate.  Using
an analysis of the CP asymmetries in SCS decays based on a $c \to u$
penguin amplitude with standard model weak phase but enhanced by
CP-conserving strong interactions, we estimate typical shifts in
$\gamma$ of several degrees and pinpoint measurements which would reduce
uncertainties to an acceptable level.
\end{quote}

\leftline{PACS numbers: 13.25.Ft, 13.25.Hw, 14.40.Lb, 14.40.Nd}
\bigskip

\section{Introduction}

The precise determination of phases of the Cabibbo-Kobayashi-Maskawa (CKM)
matrix is crucial to the understanding of CP violation.  At present there
appears to be reasonable agreement on magnitudes and phases of CKM matrix
elements \cite{CKMfitter,UTfit}.  However, discrepancies among different
determinations of these quantities can signal new physics, for example
due to new heavy particles entering into loop diagrams.  Consequently,
it is essential to pursue the widest variety of measurements of CKM
elements.

One quantity which is determined indirectly with reasonable accuracy but
whose direct measurement has lagged with respect to many others is the
weak phase $\gamma$, related to CKM elements $V_{ij}$ by $\gamma = {\rm Arg}
(- V^*_{ub} V_{ud}/V^*_{cb} V_{cd})$.  A promising method for measuring
$\gamma$ directly, proposed by Gronau, London, and Wyler (GLW)  \cite{GLW},
compared rates for $B^+ \to D K^+$ and $B^- \to D K^-$, where $D$ is a
superposition of $D^0$ and $\od$ decaying to a CP eigenstate.  In the initial
formulation of this method, CP violation in charm decays was assumed
negligible, as suggested by standard model (SM)
estimates~\cite{Buccella:1994nf}.

Variants of the GLW method include $B^\pm \to D K^{*\pm}, D^*(\to D\pi^0,
D\gamma) K^\pm$ and $B^0 \to D K^{*0}$ where $D$ decays to CP eigenstates, and
processes of this kind in which $D^0$ and $\bar D^0$ decay to a common flavor
state such as $K^-\pi^+$ \cite{ADS} or to a three-body self-conjugate final
state such as $K_S \pi^+\pi^-$~\cite{GGSZ}.  Results obtained in these
processes have been reported by the
{\sc BaBar} \cite{delAmoSanchez:2010ji,Lees:2013zd}, Belle
\cite{Abe:2006hc,Trabelsi:2013uj},
CDF~\cite{Aaltonen:2009hz}, and LHCb
\cite{Aaij:2012kz,LHCbCONF,Malde:2013kf,LHCbK*0} collaborations.

A value of several tenths of a percent has now been seen for
$\Delta A_{CP}$, the difference between the CP asymmetries of $D^0 \to K^+ K^-$
and $D^0 \to \pi^+ \pi^-$ \cite{LHCbasym,CDFasym}.  These two asymmetries have
been included in a recent experimental study by LHCb of $B^\pm \to D K^\pm$
\cite{LHCbCONF}, concluding that their effect on determining
$\gamma$  is marginal at the current level of experimental precision.
Refs.\ \cite{Wang:2012ie,MZ} have shown that unless CP violation in $D$ decays
is taken into account, the determination of $\gamma$ via the GLW
method can be shifted from its true value by up to several degrees.

Three of us have previously assumed that CP violation in charm decays is due to
a penguin amplitude with the SM phase but enhanced by CP-conserving
strong-interaction effects \cite{BGR,Bhattacharya:2012kq,Bhattacharya:2012pc}.
The possibility of such an enhancement was pointed out some time ago, in
analogy to the likely enhancement of penguin amplitudes in $K \to 2 \pi$ decays
\cite{GG,Sav}.  A large number of authors have suggested that asymmetries at
the level observed in $\Delta A_{CP}$ cannot be excluded within the CKM
framework \cite{Bigi:2011em}.  A consistent description of SCS CP-violating
charm decays was found in Refs.\,\cite{BGR,Bhattacharya:2012kq,Bhattacharya:%
2012pc}, and predictions were made for correlations between CP asymmetries in
several decays of charmed mesons to pairs of light pseudoscalar mesons $P$.

In the present paper we apply this description to $B^\pm \to D K^\pm, D \to
\pi^+\pi^-, K^+ K^-$, studying within the CKM framework shifts in $\gamma$ due 
to CP violation in $D \to PP$ decays.  We find shifts of up to a few degrees as noted in Refs.\ \cite{Wang:2012ie,MZ}, where no specific scheme has been used
for CP violation and no correlations existed between CP asymmetries in $D$
decays.  We identify the most crucial measurements for reducing
the uncertainty in effects of these shifts to acceptable levels, i.e., below
effects of other uncertainties.

We review the GLW method in Sec.\ \ref{sec:GLW} allowing for CP violation 
in charm decays, using observables involving minimal systematic uncertainties.
Present information on direct CP asymmetries in SCS $D$ decays to two
pseudoscalars is
summarized in Sec.\ \ref{sec:Adirs}.  The approach of Ref.\ \cite{BGR} is then
outlined in Sec.\ \ref{sec:BGR}, and applied to obtain predictions for shifts
in $\gamma$ in Sec.\ \ref{sec:shifts}.  We summarize in Sec.\ \ref{sec:concl}.

\section{GLW method in presence of charm CP violation \label{sec:GLW}}

We shall be concerned with a single source of information on $\gamma$: the
decays $B^\pm \to D K^\pm$, with $D$ decaying to a CP-even eigenstate
such as $f_D = \pi^+\pi^-$ or $K^+K^-$.  Our notation will follow that of Ref.\
\cite{MZ}.  We define
\bea
A(B^-\to D^0 K^-) &\equiv& A_B \nn\\
A(B^-\to\od K^-) &\equiv& A_B r_B e^{\i(\delta_B - \gamma)} \\
A(D^0\to f_D) &\equiv& A_f \nn~,
\eea
where we have taken a strong phase to be zero in $B^- \to
D^0 K^-$ with no loss of generality.  By CP conjugation we then have
\bea
A(B^+\to \od K^+) &\equiv& A_B \nn\\
A(B^+\to D^0 K^+) &\equiv& A_B r_B e^{\i(\delta_B + \gamma)} \\
A(\od \to f_D) &\equiv& \bA_f \nn~.
\eea

The parameters $r_B$ and $\delta_B$ are measurable by combining information
from $B^\pm \to D K^\pm$, where neutral $D$ mesons decay to CP-eigenstates,
flavor-specific states or $K_S\pi^+\pi^-$. Current values taken from Ref.\
\cite{CKMfitter}
are:
\beq \label{eqn:Bpar}
r_B = 0.099 \pm 0.008~,~~\delta_B = (110\pm15)^\circ~.
\eeq

The magnitudes of the amplitudes $|A_f|$ and $|\bA_f|$ are measurable
through the CP-averaged branching ratio for $D^0 \to f$ and $\od \to
f$ (giving $|A_f|^2 + |\bA_f|^2$) and the direct CP asymmetry
\beq
\label{AdirCP}
A_{CP}^{\rm dir}(f) = \frac{|A_f|^2 - |\bA_f|^2}{|A_f|^2 + |\bA_f|^2}~.
\eeq
We define the weak
(CP-violating) phase $\alpha_f \equiv {\rm Arg}(A_f/\bA_f)$. Its value is, as
yet, unspecified.  If $\phi$ is taken as the phase of $A_f$ we have
\beq
A_f = |A_f| e^{\i\phi}~;~~\bA_f = |\bA_f| e^{\i(\phi-\alpha_f)}~.
\eeq

Using the above definitions we may now construct the following amplitudes:
\bea
A(B^-\to f_D K^-) &=& A_B~A_f + \bA_f~A_B~r_B~e^{\i(\delta_B - \gamma)} \nn \\
&=& A_B~\(|A_f| + |\bA_f|~r_B~e^{\i(\delta_B - \gamma - \alpha_f)}\)~e^{\i\phi} ~,
\label{Eq:mi}\\
A(B^+\to f_D K^+) &=& A_B~\bA_f + A_f~A_B~r_B~e^{\i(\delta_B + \gamma)} \nn \\
&=& A_B~\(|\bA_f| + |A_f|~r_B~e^{\i(\delta_B + \gamma + \alpha_f)}\)~e^{\i(\phi
- \alpha_f)} ~. \label{Eq:pl}
\eea
The squared magnitudes of Eqs.\ (\ref{Eq:mi}) and
(\ref{Eq:pl}) give \cite{MZ}
\bea\label{Eq:magmi}
|A(B^-\to f_DK^-)|^2&=&|A_B|^2~\(|A_f|^2 + r^2_B
~|\bA_f|^2\right. \nn \\ && \left. + 2~r_B~|A_f||\bA_f|\cos(\delta_B - \gamma - \alpha_f)\)~,
\eea
\bea\label{Eq:magpl}
|A(B^+\to f_DK^+)|^2&=&|A_B|^2~\(|\bA_f|^2 + r^2_B
~|A_f|^2 \right. \nn \\ && \left. + 2~r_B~|A_f||\bA_f|\cos(\delta_B + \gamma + \alpha_f)\)~.
\eea
Adding and subtracting the above equations we may form quantities that
are relevant in constructing the GLW observables:
\bea
|A(B^-\to f_DK^-)|^2 + |A(B^+\to f_DK^+)|^2~=~|A_B|^2(|A_f|^2 + |\bA_f|^2)\nn \\
\(1 + r^2_B + 2~r_B \cos\delta_B\cos(\gamma + \alpha_f)\sqrt{1 - (A_{CP}^{\rm dir}(f))^2}\)~,
\\
|A(B^-\to f_DK^-)|^2 - |A(B^+\to f_DK^+)|^2~=~~|A_B|^2\nn(|A_f|^2 + |\bA_f|^2) \\
\(A_{CP}^{\rm dir}(1 - r^2_B) + 2~r_B\sin\delta_B\sin(\gamma + \alpha_f)
\sqrt{1 - (A_{CP}^{\rm dir}(f))^2}\) ~.
\eea
The last expression differs from a similar one in Ref.\ \cite{MZ} by a term
$(1 - r^2_B)~(|A_f|^2 - |\bA_f|^2)$, which vanishes only in the absence of
direct CP violation in charm and hence cannot be neglected.

We now take the expressions for $A_f$ and $\bA_f$ to be \cite{MZ}
\beq\label{AmpD}
A_f = |A_f^0|(1 + r_f e^{i (\delta_f - \gamma)})~,~~
\bA_f = |A_f^0|(1 + r_f e^{i (\delta_f + \gamma)})~,
\eeq
where $A_f^0$ is the amplitude in the absence of a CP-violating term, and
we have assumed that the source of CP violation has the SM phase $-\gamma$ as
in Ref.\ \cite{BGR}.
The direct CP asymmetry [Eq.~(\ref{AdirCP})] is then given by
\beq\label{Adir}
A_{CP}^{\rm dir}(f) = \frac{2 r_f \sin \delta_f \sin \gamma}{1 + r_f^2 + 
2 r_f \cos \delta_f \cos \gamma} ~.
\eeq
The fact that  $A_{CP}^{\rm dir}(f)$ is of order a few times $10^{-3}$ in $D^0 \to
\pi^+\pi^-$ and $D^0 \to K^+ K^-$ (see Tables~\ref{tab:DACP}~and~\ref{tab:ACP}
in Section \ref{sec:Adirs}) suggests that $r_f$ in these processes is also of
this order as argued in Refs.~\cite{BGR,Bigi:2011em}.  Our subsequent analysis
(in particular the discussion in Secs. \ref{sec:BGR} and \ref{sec:shifts})
applies also to larger value of $r_f$, for instance of order $10^{-2}$, with an
order of magnitude further suppression provided by $\sin\delta_f$.  We ignore
the fine-tuned solution where $r_f$ is large, but the strong phases in these
decays are of order $10^{-3}$. One can show that
\bea
\label{alphaf}
\alpha_f &=&  - \tan^{-1}\(\frac{2 r_f\cos\delta_f\sin\gamma + r^2_f\sin2\gamma}
{1 + 2 r_f\cos\delta_f\cos\gamma + r^2_f\cos2\gamma}\) \nonumber \\
&\approx&- A_{CP}^{\rm dir}(f)\cot \delta_f ~,
\eea
where the last approximation holds to leading order in $r_f$.

It has been suggested in Ref.\ \cite{Gronau:2002mu} that when applying
the GLW method one normalizes the CP-averaged rate for $B^\pm \to f_DK^\pm$
by that for $B^\pm \to f_D \pi^\pm$,
\beq\label{RK/pi}
R^f_{K/\pi}  \equiv  \frac{\Gamma(B^- \to f_D K^-) + \Gamma(B^+ \to f_D K^+)}
{\Gamma(B^- \to f_D \pi^-) + \Gamma(B^+\to f_D \pi^+)}~,
\eeq
and one takes the ratio of this  fraction and a corresponding fraction for $D^0$ flavor state,
\beq
R(K/\pi) \equiv \frac{\Gamma(B^- \to D^0 K^-)}{\Gamma(B^- \to D^0 \pi^-)}~.
\eeq
Significant experimental systematic uncertainties cancel in these
fractions~\cite{delAmoSanchez:2010ji,Abe:2006hc,Aaltonen:2009hz,LHCbCONF}.
Defining ratios of amplitudes and strong phases in $B^- \to D \pi^-$ in analogy with
$B^- \to D K^-$,
\beq
\frac{A(B^- \to \bar D^0 \pi^-)}{A(B^- \to D^0 \pi^-)} \equiv
r_{B(\pi)} e^{i(\delta_{B(\pi)} - \gamma)}~,
\eeq
one may express the double fraction
\beq
R^f_{CP+} \equiv \frac{R^f_{K/\pi}}{R(K/\pi)}
\eeq
in terms of $\gamma$ and these parameters.

In the absence of CP violation in $D^0 \to f_D$, one has~\cite{Gronau:2002mu}
\beq\label{Eq:RnoCP}
R^f_{CP+} = \frac{1 + r^2_B + 2~r_B \cos\delta_B\cos\gamma}
{1 + r^2_{B(\pi)} + 2~r_{B(\pi)} \cos\delta_{B(\pi)}\cos\gamma}~,
\eeq
where the parameter $r_{B(\pi)}$ is expected to be very small,
$r_{B(\pi)} \sim r_B\tan^2\theta_C \sim 0.005$. [See Eq.\,(\ref{eqn:Bpar}).]
We note that in the approximation of neglecting CP violation in $D^0 \to f_D$ the ratio
$R^f_{K/\pi}$ may be defined as $R^f_{K/\pi} \equiv [\Gamma(B^- \to D_{CP+}K^-)
+ \Gamma(B^+ \to D_{CP+} K^+)]/
[\Gamma(B^- \to D_{CP+} \pi^-) + \Gamma(B^+\to D_{CP+} \pi^+)]$. Consequently
this ratio and the double ratio $R^f_{CP+} $ do not depend on $f_D$.

Including CP violation in $D^0 \to f_D$ one finds an expression for $R^f_{CP+}$ which
depends on $f_D$ through the CP asymmetry $A_{CP}^{\rm dir}(f)$,
\beq
\label{Eq:apR}
R^f_{CP+} = \frac{1 + r^2_B + 2~r_B \cos\delta_B\cos(\gamma + \alpha_f)
\sqrt{1 - (A_{CP}^{\rm dir}(f))^2}}{1 + r^2_{B(\pi)}
+ 2~r_{B(\pi)} \cos\delta_{B(\pi)}\cos(\gamma + \alpha_f)
\sqrt{1 - (A_{CP}^{\rm dir}(f))^2}}~.
\eeq
Neglecting corrections in $R^f_{CP+}$ which are quadratic in $A^{\rm dir}_{CP}(f)$
[${\cal O}(10^{-5})]$ and using Eq.\,(\ref{alphaf}), we note that corrections linear in
$A^{\rm dir}_{CP}(f)\sim {\rm few}\times 10^{-3}$ are multiplied by $r_B$ and are therefore
negligible relative to $R^f_{CP+} = 1 + {\cal O}(r_B)$. Thus, a comparison of
Eqs.\ (\ref{Eq:RnoCP}) and (\ref{Eq:apR}) shows that the determination of
$\gamma$ is not affected in a significant way by including $A^{\rm dir}_{CP}(f)$ in
$R^f_{CP+}$.

The other measurable quantity used in the GLW method is the CP asymmetry
$A^f_{CP+}$,
\beq
A^f_{CP+} \equiv \frac{\Gamma(B^-\to f_DK^-)-\Gamma(B^+\to f_DK^+)}
{\Gamma(B^-\to f_DK^-) + \Gamma(B^+\to f_DK^+)}~.
\eeq
In the absence of CP violation in $D^0 \to f_D$, this asymmetry may be defined
as $[\Gamma(B^-\to D_{CP+}K^-)-\Gamma(B^+\to D_{CP+}K^+)]/[\Gamma
(B^-\to D_{CP+}K^-) +  \Gamma(B^+\to D_{CP+}K^+)]$ which is independent of
$f_D$ and is given by~\cite{Gronau:2002mu}
\beq\label{ACPnodirCP}
A^f_{CP+} = \frac{2~r_B \sin\delta_B \sin\gamma}{1 + r^2_B +
2~r_B\cos\delta_B\cos\gamma}~.
\eeq
When  including CP nonconservation in $D^0 \to f_D$ the asymmetry $A^f_{CP+}$
becomes dependent on $A_{CP}^{\rm dir}(f)$. Neglecting terms quadratic in
$A_{CP}^{\rm dir}(f)$, one finds
\beq\label{eqn:ACP}
A^f_{CP+} = \frac{A_{CP}^{\rm dir}(f)(1 - r^2_B) + 2~r_B \sin\delta_B
\sin(\gamma + \alpha_f)}{1 + r^2_B + 2~r_B\cos\delta_B\cos(\gamma + \alpha_f)}~.
\eeq
The two terms in the numerator and the last term in the denominator involve corrections
 linear in $A^{\rm dir}_{CP}(f)$ modifying the expression (\ref{ACPnodirCP}) for the case of
 no direct asymmetry.

We note that an expression independent of $A^{\rm dir}_{CP}(f)$ similar to
(\ref{ACPnodirCP}), but with an opposite overall sign and an
${\cal O}(r^2_B)$ correction with opposite sign, describes the asymmetry for
Cabibbo-favored $D^0$ decays to CP-odd eigenstates (such as $K_S \phi$
and $K_S \pi^0$) where CP violation is negligible in the CKM framework,
\beq\label{ACP-}
A^f_{CP-} = -\frac{2~r_B \sin\delta_B \sin\gamma}{1 + r^2_B -
2~r_B\cos\delta_B\cos\gamma}~.
\eeq
A precise measurement of this asymmetry could avoid uncertainties from
$A^{\rm dir}_{CP}(f)$.

Writing
\beq\label{gamma-eff}
A^f_{CP+} = \frac{2~r_B \sin\delta_B \sin\gamma_{\rm eff}}{1 + r^2_B +
2~r_B\cos\delta_B\cos\gamma_{\rm eff}}~,
\eeq
we assume a measurement of $A^f_{CP+}$ from which $\gamma_{\rm eff}$ is determined.
Our purpose is then to study the shift $\delta\gamma \equiv \gamma - \gamma_{\rm eff}$
as a function of $A^{\rm dir}_{CP}(f)$.
Defining  $\gamma + \alpha_f = \gef + \delta\gamma + \alpha_f \equiv \gef + x$,
we expand the numerator and denominator in (\ref{eqn:ACP})  to first order in
$x$.  Comparing this expression with (\ref{gamma-eff}), cross-multiplying the
two ratios, cancelling the leading terms and keeping terms linear in $x$ and
$A_{CP}^{\rm dir}(f)$, one finds
\bea \label{eqn:dg}
\delta \gamma & = &  - \alpha_f - A_{CP}^{\rm dir}(f) \left[\frac{1-r_B^2}{2  r_B \sin\delta_B} \,
\frac{1 + r_B^2 + 2  r_B \cos\delta_B
\cos\gef}{(1 + r_B^2) \cos\gef + 2  r_B \cos\delta_B} \right] \nonumber\\
& \approx & A_{CP}^{\rm dir}(f) \left[ \cot \delta_f -
\frac{1-r_B^2}{2  r_B \sin\delta_B} \, \frac{1 + r_B^2 + 2  r_B \cos\delta_B
\cos\gef}{(1 + r_B^2) \cos\gef + 2  r_B \cos\delta_B} \right] \nonumber \\
& \approx & 2r_f\cos\delta_f\sin\gef -A_{CP}^{\rm dir}(f) \left[\frac{1-r_B^2}{2  r_B \sin\delta_B} \,
\frac{1 + r_B^2 + 2  r_B \cos\delta_B
\cos\gef}{(1 + r_B^2) \cos\gef + 2  r_B \cos\delta_B} \right]~.
\eea
While both corrections are exhibited as proportional to $A_{CP}^{\rm dir}(f)$,
the first term appears to diverge as $\delta_f \to 0$.  Recalling that this
correction is actually $\approx 2r_f \cos \delta_f\sin\gef$ [Eq.~(\ref{Adir})],
which is finite at $\delta_f = 0$, we will use the last expression in Sec.\
\ref{sec:shifts}.  The second term in (\ref{eqn:dg}) illustrates an
enhancement by $1/2r_B$ of the shift in $\gamma$ due to
$A^{\rm dir}_{CP}(f)$~\cite{Wang:2012ie,MZ}.

We mention in passing two approximations which have been applied to the CP
asymmetry in (\ref{eqn:ACP}).  Neglecting $\alpha_f$ [which is of the same
order in $r_f$ as $A_{CP}^{\rm dir}(f)$, but leads to a correction linear in
$r_B$] and approximating the overall coefficient of $A_{CP}^{\rm dir}(f)$ by
one, Eq. (\ref{eqn:ACP}) reduces to an expression employed in Ref.\
\cite{LHCbCONF},
\beq \label{eqn:dga}
A^f_{CP+} \approx  \frac{2 r_B \sin\delta_B \sin \gamma}
 {1 + r_B^2 + 2 r_B \cos \delta_B \cos\gamma} + A_{CP}^{\rm dir}(f)~.
\eeq
A further approximation, keeping only ${\cal O}(r_B)$ in the first term, leads
to \cite{MZ}
\beq
A^f_{CP+} \simeq 2 r_B \sin\delta_B \sin \gamma +  A_{CP}^{\rm dir}(f)~.
\eeq

A CP asymmetry in $B^\pm \to f_D K^\pm$ has been measured in Refs.\
\cite{delAmoSanchez:2010ji,Aaltonen:2009hz,Aaij:2012kz} consistent with
an estimate $A^f_{CP+}(B^\pm \to f_D K^\pm) \simeq 2r_B\sin\delta_B\sin\gamma
\sim 0.15 - 0.20$.
The current error in the world-averaged value~\cite{HFAG}, $A_{CP+} = 0.19
\pm 0.03$, is still too large for sensitivity to $A_{CP}^{\rm dir}(f)$.
The corresponding error in the world-averaged measurement, $A_{CP-} = -0.11 \pm 0.05$
favoring an opposite sign as anticipated,  is somewhat larger.

A much smaller asymmetry is expected in $B^\pm \to D
\pi^\pm$ where we estimated $r_{B(\pi)} \sim 0.005$. This asymmetry is given by
an expression similar to (\ref{eqn:ACP}),
\bea
A^f_{CP+}(B \to f_D \pi) &  = & \frac{A_{CP}^{\rm dir}(f)(1 - r^2_{B(\pi)}) +
2~r_{B(\pi)} \sin\delta_{B(\pi)}\sin(\gamma + \alpha_f)}
{1 + r^2_{B(\pi)} + 2~r_{B(\pi)}\cos\delta_{B(\pi)}\cos(\gamma + \alpha_f)}
\nonumber\\
& \approx &  2~r_{B(\pi)} \sin\delta_{B(\pi)}\sin\gamma + A_{CP}^{\rm dir}(f)~.
\eea
The two contributions, which in principle may be disentangled by measuring also
$A_{CP-}(B \to f_D \pi) \approx -2~r_{B(\pi)} \sin\delta_{B(\pi)}\sin\gamma$,
are of comparable magnitudes, each less than a percent.
Thus, while the CP-averaged rate for $B^\pm \to f_D \pi^\pm$ is suitable for
normalization [see Eq.\ (\ref{RK/pi})], a useful measurement of the
corresponding asymmetry does not seem feasible in the foreseeable future.

\section{Present information on $A^{\rm dir}_{CP}$ \label{sec:Adirs}}

Dedicated measurements of the difference
$\Delta A_{CP} \equiv A^{\rm dir}_{CP}(D^0 \to K^+ K^-)
- A^{\rm dir}_{CP}(D^0 \to \pi^+ \pi^-)$
(in which many systematic errors cancel) have been performed by the LHCb
\cite{LHCbasym} and CDF \cite{CDFasym} Collaborations, while Belle has
combined independent measurements of the two asymmetries \cite{Belleasym}
to obtain a value of $\Delta A_{CP}$.  The results are shown in Table
\ref{tab:DACP}.  We shall assume that measured asymmetries are equal to direct
ones, neglecting possible contributions from indirect (mixing-induced)
asymmetries which would lead to slightly different averages
\cite{CKMfitter,HFAG}.

\begin{table}[h]
\caption{Experimental results on $\Delta A_{CP} \equiv A^{\rm dir}_{CP}(D^0 \to K^+ K^-)
- A^{\rm dir}_{CP}(D^0 \to \pi^+ \pi^-)$.
\label{tab:DACP}}
\begin{center}
\begin{tabular}{c c} \hline \hline
Reference & Value (\%) \\ \hline
LHCb \cite{LHCbasym} & $-0.82\pm0.21\pm0.11$ \\
CDF \cite{CDFasym} & $-0.62\pm0.21\pm0.10$ \\
Belle \cite{Belleasym} & $-0.87\pm0.41\pm0.06$ \\
Average & $-0.74 \pm 0.15$ \\ \hline \hline
\end{tabular}
\end{center}
\end{table}
\begin{table}[h]
\caption{Experimental results on some direct CP asymmetries in $D$ decays.
\label{tab:ACP}}
\begin{center}
\begin{tabular}{c c c} \hline \hline
Decay & Reference & Value (\%) \\ \hline
$D^0 \to \pi^+\pi^-$ & CDF \cite{CDFacp} & $0.22\pm0.24\pm0.11$ \\
 & Belle \cite{Belleasym} & $0.55\pm0.36\pm0.09$ \\
 & Average & $0.33\pm0.22$ \\
$D^0 \to K^+K^-$ & CDF \cite{CDFacp} & $-0.24\pm0.22\pm0.09$ \\
 & Belle \cite{Belleasym} & $-0.32\pm0.21\pm0.09$ \\
 & Average & $-0.28\pm0.16$ \\
$D^+ \to K^+ \ok$ (a) & {\sc BaBar} \cite{Lees:2012jv} & $0.46\pm0.36\pm0.25$\\
 & Belle \cite{Belleasym,Ko:2012uh} & $0.08\pm0.28\pm0.14$ \\
 & Average & $0.21\pm0.25$ \\ \hline \hline
\end{tabular}
\end{center}
\begin{center}
(a) After subtraction of CP asymmetry due to $\ko$--$\ok$ mixing.
\end{center}
\end{table}

The average in Table \ref{tab:DACP} will be used in the next Section to
constrain the magnitude of a SM penguin amplitude as a function of its
strong phase.  (Slightly different averages were used in Refs.\ \cite{Wang} and
\cite{MZ}.) In addition weak constraints on this strong phase will be
seen to result from measured CP asymmetries in individual final states,
quoted in Table \ref{tab:ACP}.

\section{Charm CP violation with enhanced SM penguin \label{sec:BGR}}

In Ref.\ \cite{BGR} three of us have calculated CP asymmetries for
several $D \to PP$ decays, where $P=\pi,K$, assuming that the nonzero
value of $A_{CP}^{\rm dir}$ is due to a SM penguin amplitude with the weak phase of the
standard model $c \to b \to u$ loop diagram, but with a CP-conserving
enhancement as if due to the strong interactions.
In this case the magnitude
and strong phase of this amplitude $P_b$ are correlated in order to fit the
observed CP asymmetry, allowing the prediction of CP asymmetries for other
singly-Cabibbo-suppressed modes.  We refer the reader to that work for
details, but outline the method briefly here.

For the decay of a charmed meson $D$ to any final state $f$ we are defining
a direct CP asymmetry using the same convention as in (\ref{AdirCP})
\beq
A_{CP}^{\rm dir}(f) \equiv \frac{\Gamma(D \to f) - \Gamma(\bar D \to \bar f)}
                      {\Gamma(D \to f) + \Gamma(\bar D \to \bar f)}~.
\eeq
We take the CP-conserving amplitudes from a flavor-SU(3)
description of charm decays
presented previously \cite{Bhattacharya:2008ke,Bhattacharya:2009ps}.  A new
ingredient \cite{BGR} with respect to that work is a U-spin breaking $c \to u$
penguin amplitude whose $V^*_{cd(s)}V_{ud(s)}$ term reproduces satisfactorily
decay rates for singly-Cabibbo-suppressed (SCS) processes, while its
$V^*_{cb}V_{ub}$ term accounts for $\Delta A_{CP}$. This scheme allows the
prediction of other CP asymmetries in SCS charmed meson decays.

Denoting by $\delta_f$ a phase defined in Ref.\ \cite{BGR} as $\phi^f$, we
write the amplitude for a decay $D \to f$
in a manner similar to (\ref{AmpD})
\beq
{\cal A}(D \to f) \equiv A_f = |T_f| e^{i \phi^f_T} (1 + r_f e^{i(\delta_f-\gamma)})~.
\eeq
Here $T_f$ represents terms with the weak phase of the tree-level terms
contributing to that amplitude, $\phi_T^f$ is its strong phase, $r_f$ is
the ratio of the magnitude of the CP-violating penguin contribution to that of
$T_f$, $-\gamma$ is the weak phase of the CP-violating penguin, and $\delta_f$
is the strong phase of the CP-violating penguin relative to $T_f$.  Here one
has
\beq
\delta_f ={\rm Arg}(P_b) - \phi_T^f + \gamma~,~~P_b = p e^{i(\delta-\gamma)}~,
\eeq
leading to the relation $\delta_f = \delta - \phi_T^f$.
The magnitudes $T_f$ and phases $\phi_T^f$ for some $D \to PP$ processes
\cite{BGR} are summarized in Table \ref{tab:Tphi}.  For $D^0 \to \pi^0 \pi^0$,
the amplitude must be multiplied by an additional factor of $-1/\sqrt{2}$
\cite{BGR}.

\begin{table}
\caption{Magnitudes $T_f$ and phases $\phi_T^f$ for some $D \to PP$ processes.
\label{tab:Tphi}}
\begin{center}
\begin{tabular}{c c c} \hline \hline
Decay &     $|T_f|$     & $\phi_T^f = {\rm Arg}(T_f)$ \\
mode  & ($10^{-7}$ GeV) &        (degrees) \\ \hline
$D^0 \to \pi^+ \pi^-$ & 4.70 & --158.5 \\
$D^0 \to K^+ K^-$     & 8.48 &    32.5 \\
$D^0 \to \pi^0 \pi^0$ & 3.51 &    60.0 \\
$D^+ \to K^+ \ok$     & 6.87 &   --4.2 \\ \hline \hline
\end{tabular}
\end{center}
\end{table}

\begin{figure}
\begin{center}
\includegraphics[width=0.52\textwidth]{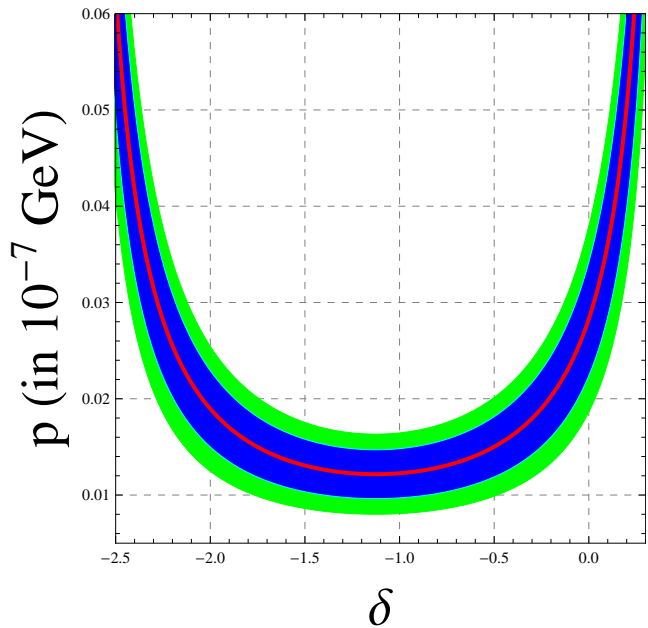}
\end{center}
\caption{Magnitude $p$ of the CP-violating penguin amplitude as a function
of its strong phase $\delta$.  The red curve was obtained using the value
$\Delta A_{CP} = -0.74\%$, while the inner (blue) and outer (green) bands
respectively correspond to $\pm 1 \sigma$ and $\pm 1.64 \sigma$ shifts from
this value, where $\sigma(\Delta A_{CP}) = 0.15\%$.  The plot is shown only for
$\gamma = 67.2^\circ$, as $p$ is not very sensitive to the exact value
of $\gamma$.
\label{fig:p-delta}}
\end{figure}

\begin{figure}
\begin{center}
\includegraphics[width=0.451\textwidth]{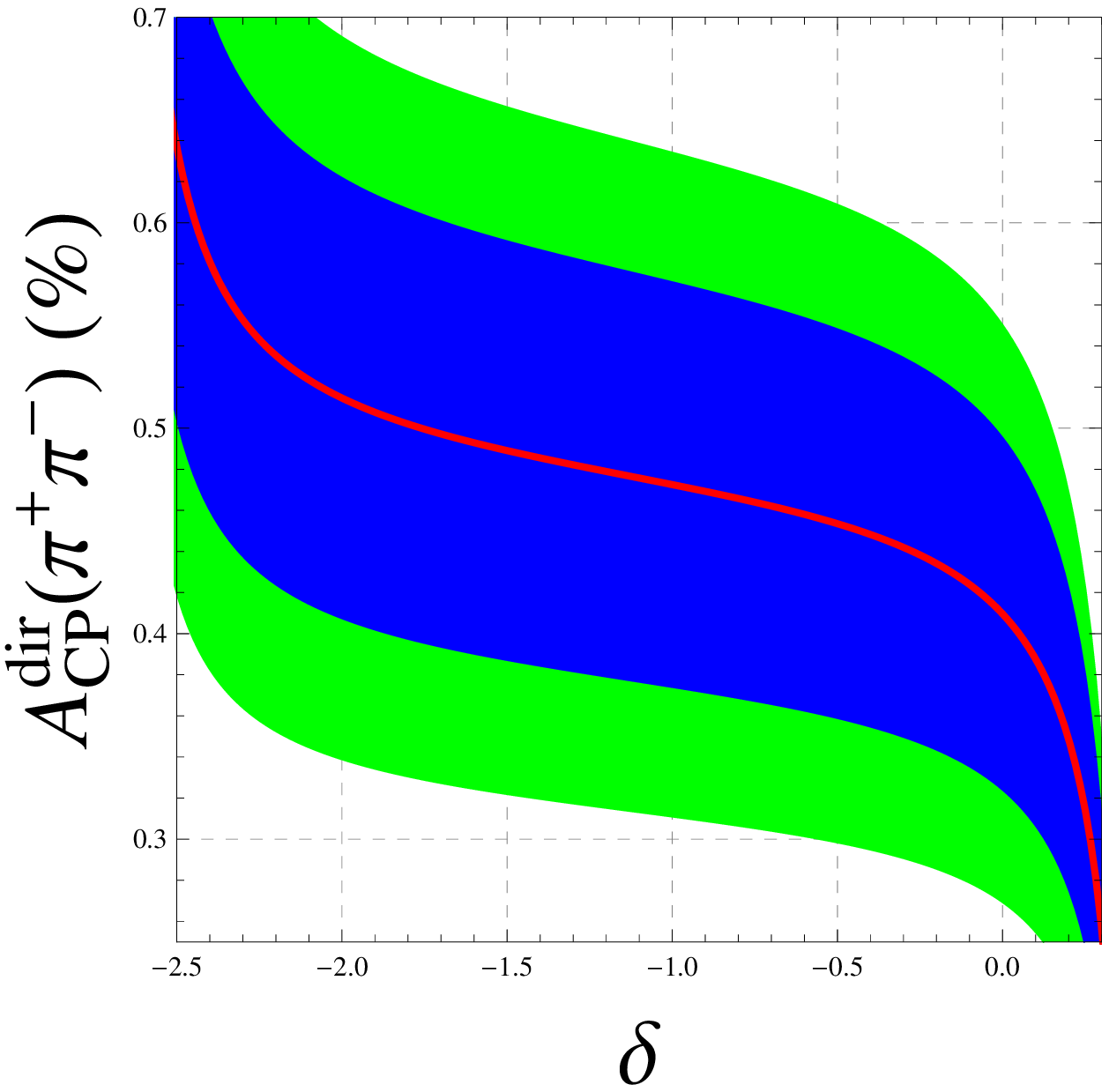} \hspace{0.3in}
\includegraphics[width=0.458\textwidth]{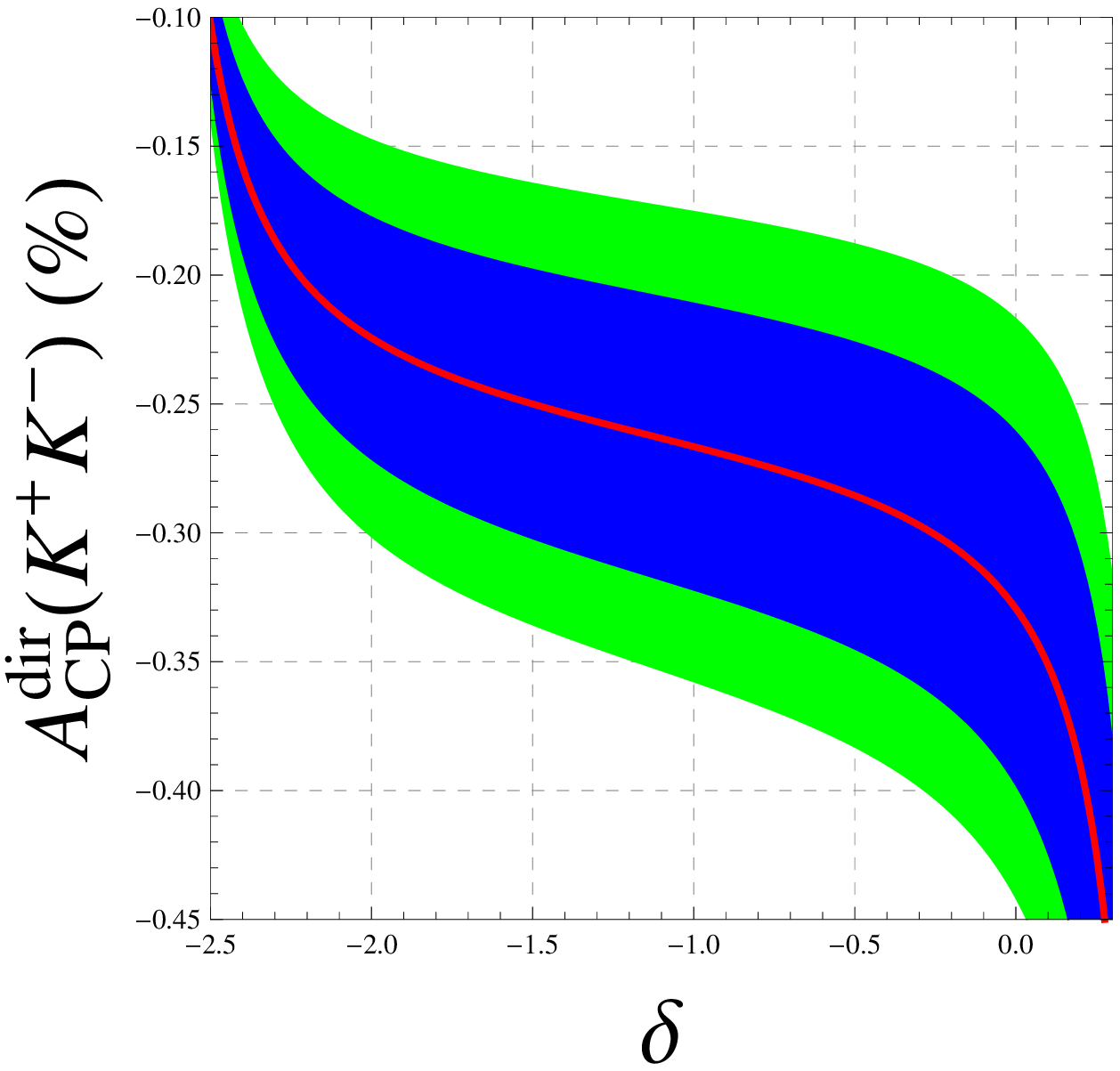}
\vskip 0.2in
\includegraphics[width=0.4545\textwidth]{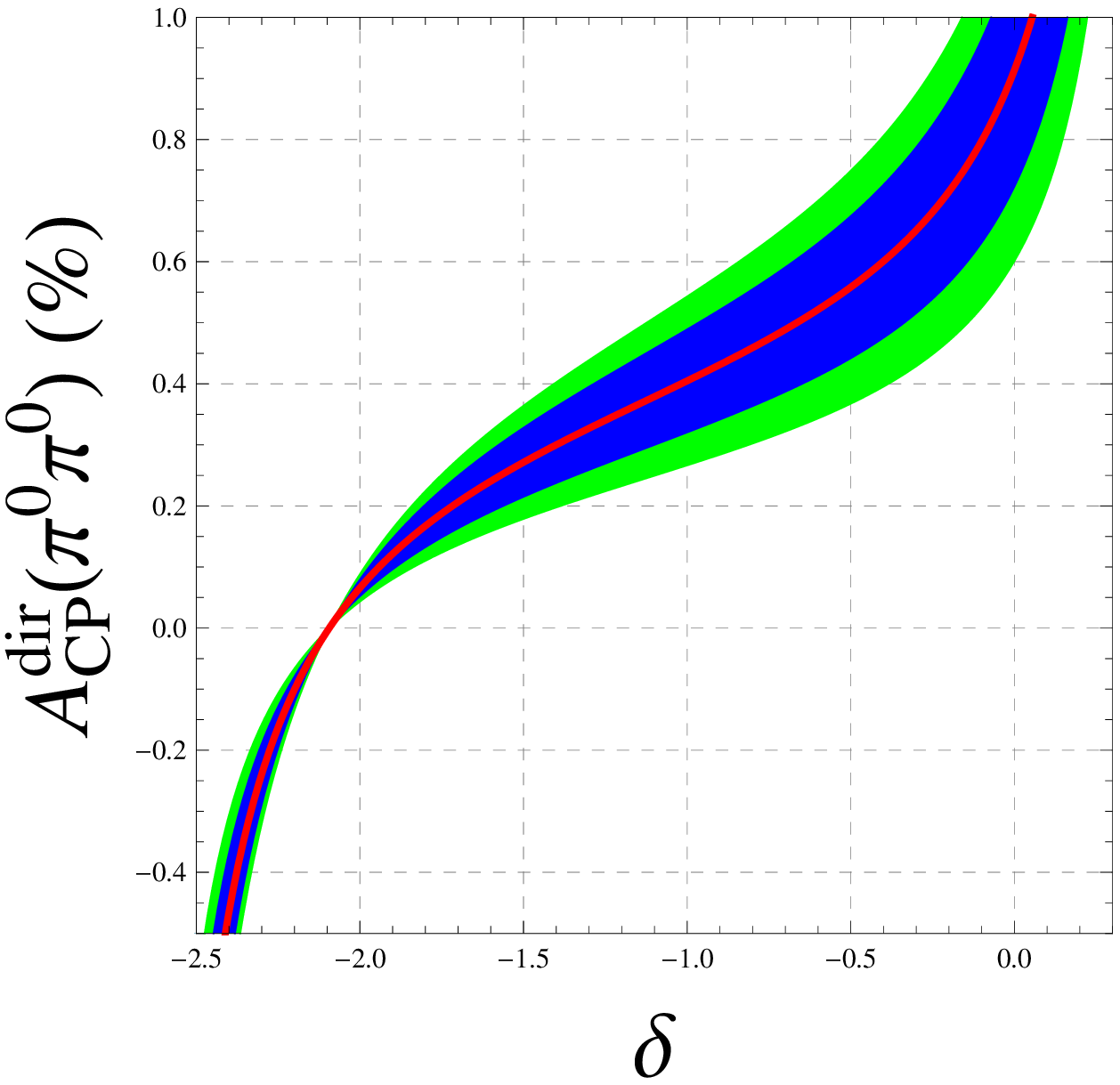} \hspace{0.3in}
\includegraphics[width=0.4545\textwidth]{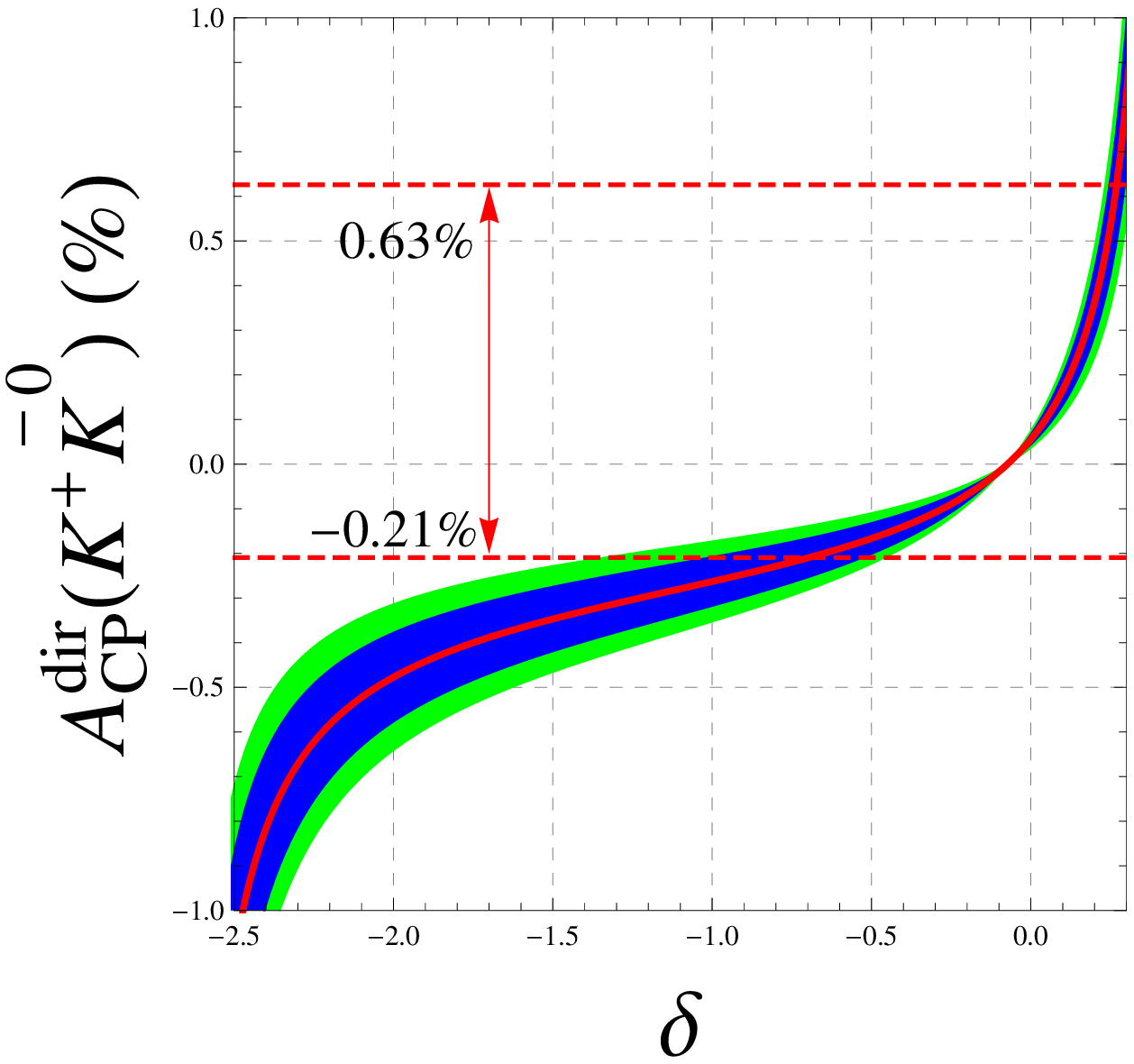}
\end{center}

\caption{Direct CP asymmetries for some SCS $D \to PP'$ decays. 
Curves and bands as in Fig.\ \ref{fig:p-delta}.  $\gamma = 67.2^\circ$ is assumed.  Very similar
plots (not shown) are obtained for $\gamma = 71.6^\circ$ and $62.6^\circ$. The dashed horizontal
(red) lines in the lower right panel denote 90\% C. L. limits based on the average of Belle
\cite{Belleasym,Ko:2012uh} and BaBar \cite{Lees:2012jv} results.
\label{fig:acp4c}}
\end{figure}

The CP asymmetries are then
\bea\label{Acp}
A_{CP}^{\rm dir}(f) &=& \frac{2\,r_f\,\sin \gamma \sin \delta_f}
{1 + r^2_f + 2\,r_f\,\cos \gamma \cos \delta_f} \nonumber\\
&=& \frac{2\,p\,|T_f|\,\sin\gamma\sin(\delta - \phi^f_T)}{|T_f|^2 + p^2 +
2\,p\,|T_f|\, \cos\gamma\cos(\delta - \phi^f_T)},
\eea
Taking $\gamma = (67.2^{+4.4}_{-4.6})^\circ$ from Ref.\ \cite{CKMfitter}, the
world-averaged asymmetry $\Delta A_{CP} = (-0.74 \pm 0.15)\%$ from Table
\ref{tab:DACP} is used to constrain the magnitude $p$ of the penguin amplitude
as a function of
its strong phase $\delta$. The value of $p$, plotted in Fig.\ \ref{fig:p-delta},
is nearly constant at several tenths of a percent of the amplitudes in Table
\ref{tab:Tphi} for a wide range of $\delta$.  This corresponds to values of
$r_f$ of this order. The value of $p$ and corresponding values of $r_f$ are
about an order of magnitude larger for extreme values of $\delta$
in Fig.\ \ref{fig:p-delta}.
The constraint on $p$ as a function of $\delta$ allows one to predict
asymmetries for (e.g.) $D^0 \to \pi^+\pi^-, K^+ K^-, \pi^0\pi^0$ and $D^+ \to
K^+ \bar K^0$, as plotted in Fig.\ \ref{fig:acp4c} for $\gamma = 67.2^\circ$.
Very similar results (not shown) are found for $\gamma = 71.6^\circ$ and
$62.6^\circ$.

For much of the range of $\delta$, $A_{CP}^{\rm dir}(\pi^+\pi^-)$ is predicted to
be positive while $A_{CP}^{\rm dir}(K^+K^-)$ is predicted to be negative.  (In
the U-spin limit they would be equal and opposite.)  This is consistent
with the central values in Table \ref{tab:ACP}.  However, the predicted
central values for $A_{CP}^{\rm dir}(K^+ \ok)$ are negative for much of the
range of $\delta$, whereas the $2 \sigma$ lower limit $A_{CP}^{\rm dir}(K^+ \ok)
> -0.3$ would tend to favor values of $\delta > -\pi/2$.  Improved
measurements of all these individual CP asymmetries would of course be
highly desirable.

\section{Predictions for shifts in weak phase $\gamma$ \label{sec:shifts}}

\begin{figure}
\begin{center}
\includegraphics[width=0.405\textwidth]{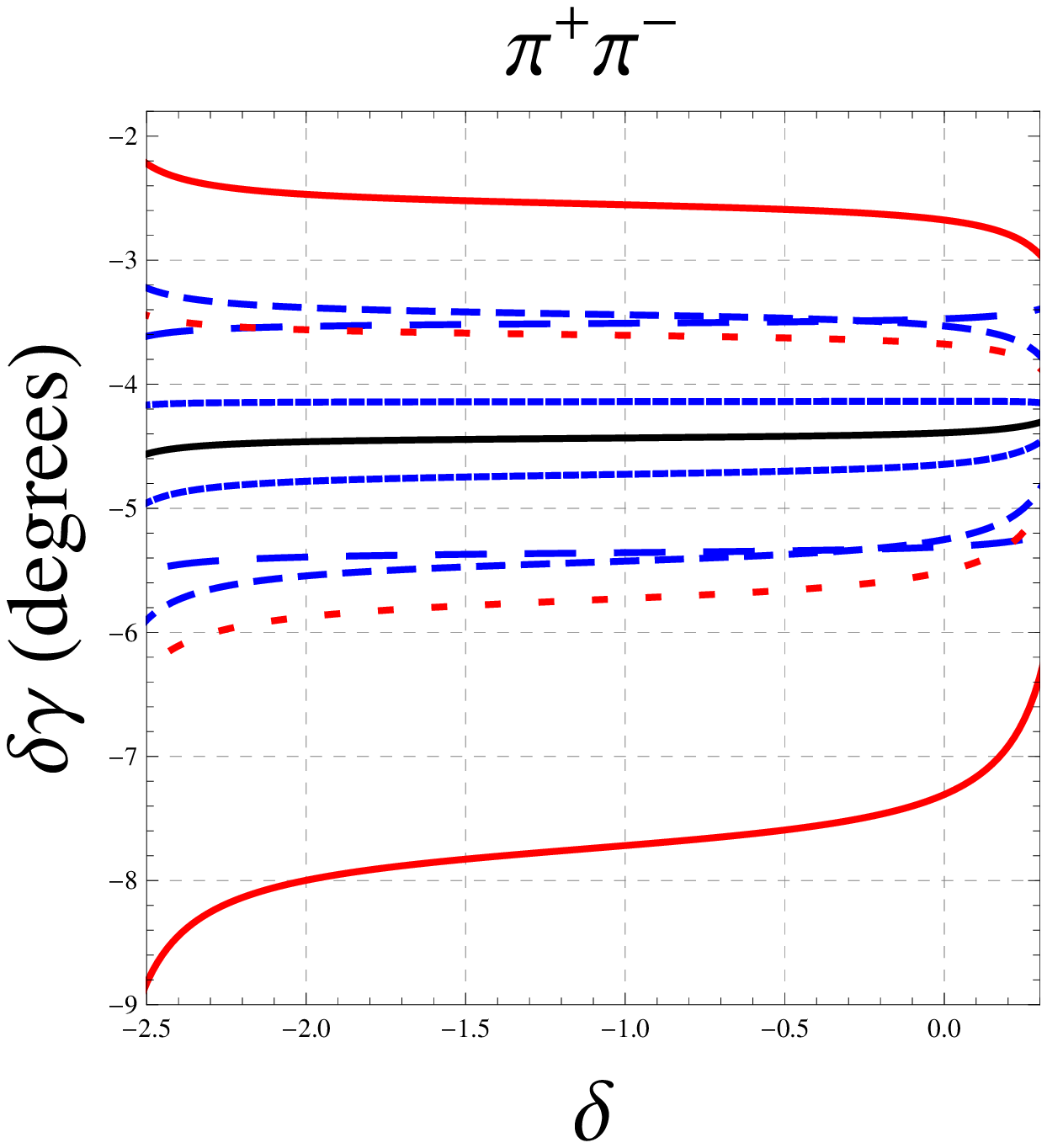} \hspace{0.3in}
\includegraphics[width=0.392\textwidth]{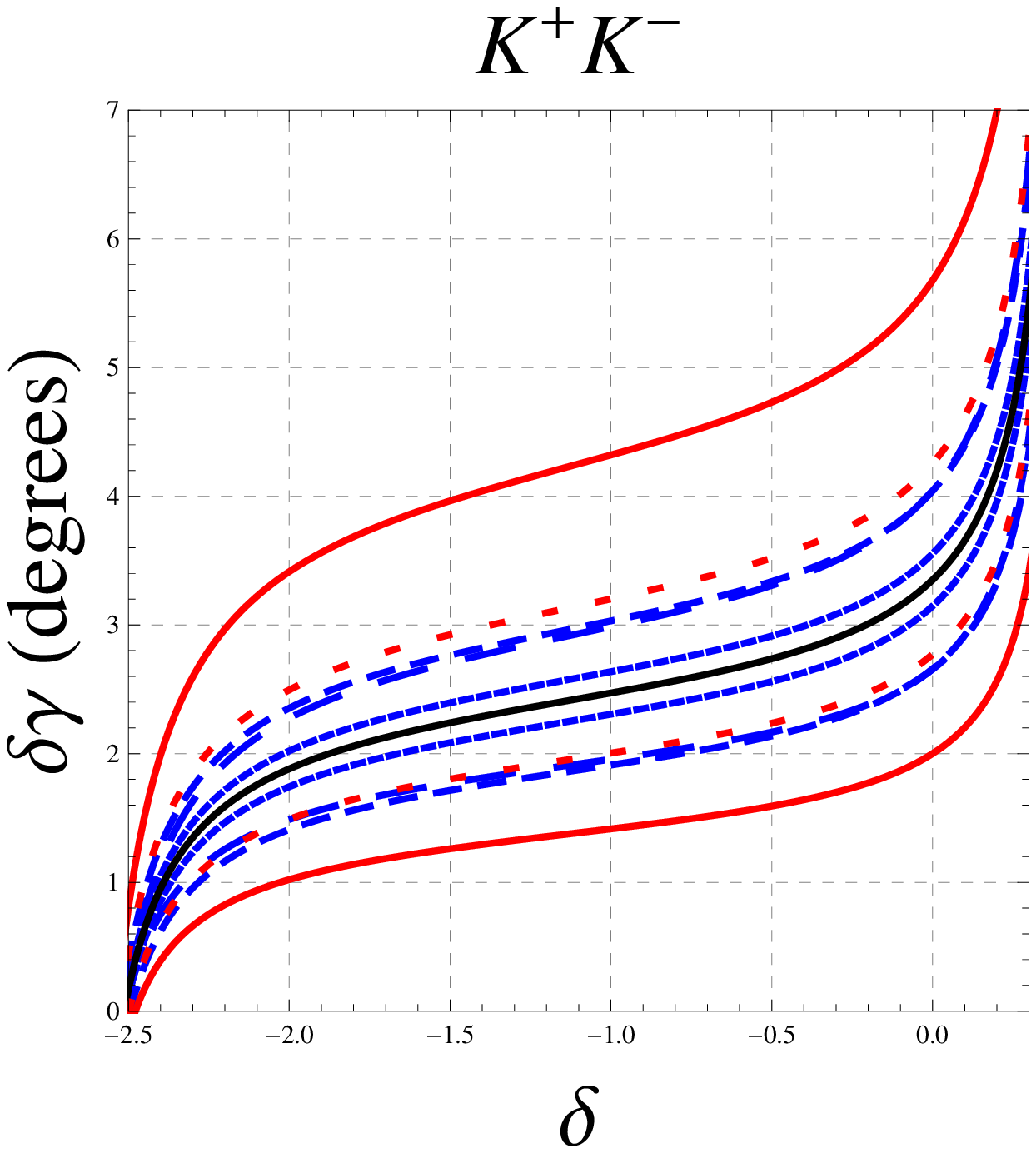}
\end{center}
\caption{Shifts $\delta \gamma$ as calculated from Eq.\ (\ref{eqn:dg}) for
$D^0\to\pi^+\pi^-$ (left panel) and $D^0\to K^+K^-$ (right panel).  The solid
central (black) curves denote the central value for $\delta\gamma$. Also shown
are $\pm 1 \sigma$ errors in $r_B$ (short dashes [blue]), $\delta_B$ (medium
dashes [blue]), and $\Delta A_{CP}$ (long dashes [blue]). The (red) dots
represent the effect of $\pm 1 \sigma$ shift in the measured value of $\gamma$
on the central (black) curve. The solid outer (red) curves denote the effect
of a $\pm 1 \sigma$ shift in the measured value of $\gamma$ on the curve that
is obtained by adding in quadrature the errors in $r_B$, $\delta_B$, and
$\Delta A_{CP}$. Here $\gamma = (67.2^{+4.4}_{-4.6})^\circ$.
\label{fig:dgamc}}
\end{figure}

Taking the predicted values of $A_{CP}^{\rm dir}$, Eq.\ (\ref{eqn:dg}) implies a shift
in the weak phase $\gamma$ associated with the use of each $D$-decay process
with a CP-even final state. In Fig.\ \ref{fig:dgamc} we present these shifts
for the final states $\pi^+\pi^-$ and $K^+K^-$. In addition to the central
value of shifts $\delta\gamma$ we also present
errors in $\delta\gamma$ due to the
variation of the various measurable parameters. In our calculations of the
shifts $\delta\gamma$ we have used the following values of $r_B$, $\delta_B$,
and $\gamma$ taken from Ref.\ \cite{CKMfitter}:
\beq\label{eqn:pars}
r_B = 0.099 \pm 0.008~,~~\delta_B = (110\pm15)^\circ~,~~\gamma = (67.2^{+4.4}%
_{-4.6})^\circ~.
\eeq

In Fig.\ \ref{fig:dgamc} the central black curves represent $\delta\gamma$
as obtained from Eq.\ (\ref{eqn:dg}) using central values of the measurable
parameters $r_B$, $\delta_B$, $\Delta A_{CP}$ and $\gamma$. The errors in
$\delta\gamma$ from the $1\sigma$ error in the measurement of the first three
parameters are shown in blue using short, medium, and long dashes
respectively. In order to obtain the effect of varying $\gamma$ on the shifts
$\delta\gamma$ we use Eq.\ (\ref{eqn:dg}) with the value of $\gamma$ set to
its $\pm 1\sigma$ limits given in Eq.\ (\ref{eqn:pars}), while the other
parameters are held fixed at their respective central values. This effect is
represented by the red dots in Fig.\ \ref{fig:dgamc}. In order obtain the
overall error in $\delta\gamma$ we first add in quadrature the errors due to
$r_B$, $\delta_B$, and $\gamma$. We then estimate the effect of varying
$\gamma$ between its $\pm 1\sigma$ limits on the quadrature sum. This effect
is represented by the solid red curves in Fig.\ \ref{fig:dgamc}.

We see in Fig.\ \ref{fig:dgamc} that the central value of the shift in
$\gamma$ for the $\pi^+\pi^-$ state is nearly constant over the entire range
of allowed values of the strong phase $\delta$. This appears to be the effect
of an accidental cancellation between a variation of $A_{CP}^{\rm dir}$ which
is modest to begin with and a compensating factor due to the variation in
$\delta_f$ for $\gamma = 67.2^\circ$. When the various sources of errors on
$\delta \gamma$ are included, this no longer holds, as depicted by the
$1\sigma$ boundary red curves in Fig.\ \ref{fig:dgamc}. On the other hand,
$\delta\gamma$ obtained using the $K^+K^-$ state shows appreciable variation
($\sim7^\circ$) over the allowed range of $\delta$. A similar exercise when
performed using the $\pi^0 \pi^0$ state yields an even larger variation in
$\delta\gamma$. However, rate asymmetries involving multiple neutral pions in
the final state are not expected to be measured with adequate precision in the
foreseeable future. We have therefore chosen to omit the $\pi^0\pi^0$ decay
mode from the present discussion.

In view of the variety of dependences of direct CP asymmetries on $\delta$, and
different behavior of shifts in $\gamma$ for $\pi^+ \pi^-$ and $K^+ K^-$ final
states, it would be beneficial to further pin down $\delta$ [e.g., with a
better measurement of $A_{CP}^{\rm dir}(K^+ \ok)$], and to apply the GLW
determination of $\gamma$ to the widest assortment of CP eigenstates in
neutral $D$ meson decays.  After the appropriate shifts $\delta \gamma$ have
been taken into account, inconsistencies in final values of $\gamma$ obtained
for different charm final states could point the way to effects of new physics.

\section{Discussion and summary \label{sec:concl}}

The determination of the weak phase $\gamma$ by means of the CP asymmetry
in $B^\pm \to D K^\pm$, followed by the decay of the neutral $D$ meson to a
CP eigenstate \cite{GLW}, must take account of asymmetries in the decays of
the neutral $D$ mesons \cite{LHCbCONF,Wang:2012ie,MZ}.  We have calculated
the corresponding shifts $\delta \gamma$ in an approach which imagines these
CP asymmetries as due to a $c \to u$ penguin amplitude with weak phase of
the standard model but enhanced by (presumably nonperturbative) strong
interaction effects beyond those anticipated by the majority of authors.
The observed value $\Delta A_{CP} \equiv A^{\rm dir}_{CP}(D^0 \to K^+ K^-) -
A^{\rm dir}_{CP}(D^0 \to \pi^+ \pi^-) = (-0.74 \pm 0.15)\%$ has been taken as a
constraint, leading to a correlation between the magnitude $p$ of the
CP-violating penguin and its strong phase $\delta$.

For $\gamma = 67.2^\circ$ the central value of the shift associated with the
decay $D^0\to\pi^+ \pi^-$ is approximately $\delta \gamma (\pi^+ \pi^-) \simeq
-4.4^\circ$, with little dependence on the allowed range of the strong phase
$\delta$.  However,
the factor multiplying $A^{\rm dir}_{CP}(f)/2r_B$ in the last expression for
$\delta \gamma$ in Eq.\ (\ref{eqn:dg}) is roughly
inversely proprtional to $\cos \gamma$, which is fairly small and fairly
sensitive to $\gamma$.  Thus, when $\gamma$ is varied within its currently
allowed range of about $\pm 4.5^\circ$, the value of $\delta \gamma$ varies
considerably.  It is further affected by uncertainties in $\Delta A_{CP}$,
$r_B$, and $\delta_B$, and acquires some dependence on $\delta$.

The shift associated with $D^0\to K^+K^-$ is of the other sign (as is
$A_{CP}^{\rm dir}$) and depends on both $\Delta A_{CP}$ and $\delta$ as well
as the uncertainties in $r_B$ and $\delta_B$.  Using measurements of
$A_{CP}^{\rm dir}$ for both these two decays, with the help of improved
knowledge of $A^{\rm dir}_{CP}(D^+ \to K^+ \ok)$ to pin down $\delta$, the
uncertainty in $\gamma$ due to CP violation in charm decay can be reduced to
a level where it is no longer the dominant uncertainty when applying the GLW
method to the decays $B^\pm \to D K^\pm$.

Let us be clear about the limitation of our study. The shift in $\gamma$ we
calculate using our $c \to u$ penguin amplitude model is based on measuring
$R_{CP+}$ and $A_{CP+}$ in $B^- \to(\pi^+\pi^-)_D K^-,~B^- \to (K^+K^-)_D
K^-$ and their CP conjugates, taking $r_B$ and $\delta_B$ as given.  This
assumes that one has measured $r_B$ and $\delta_B$ first and then uses only
these GLW processes to determine $\gamma$.

An actual analysis for determining $r_B$, $\delta_B$, and $\gamma$ in $B^\mp
\to D K^\mp$
\cite{Lees:2013zd,Trabelsi:2013uj,Aaltonen:2009hz,Aaij:2012kz,LHCbCONF}
combines information from $D$ decays to CP eigenstates \cite{GLW}, flavor
states \cite{ADS} and three-body
self-conjugate final states \cite{GGSZ}.  In this global
analysis $\gamma$ is also constrained by rates and asymmetries in $B \to D K$
where there is no direct CP violation in $D$ decay, for instance in decays to
flavor states and CP-odd eigenstates.  [See Eq.\ (\ref{ACP-})].
Thus any actual determination of $\gamma$ from $B \to D K$ will involve a
considerably smaller shift than we calculate.

Furthermore, it has been noted \cite{Gronau:2002mu,Isi,Gronau:2007xg} that
in the self-tagged decays $B^0 \to D K^{*0}$ the ratio $r^*_B$ of $B^0 \to D^0
K^{*0}$ and $B^0 \to \bar D^0 K^{*0}$ amplitudes, both of which are
color-suppressed, is expected to be about three times larger than $r_B$
defined in $B^+\to D K^+$.  We have seen an enhancement by $1/2r_B$ of the
shift in $\gamma$ due to $A^{\rm dir}_{CP}(\pi^+\pi^-, K^+ K^-)$ in $B^+ \to
D K^+$.  This implies that when applying the GLW method to $B^0 \to D K^{*0}$
the shift in $\gamma$ due to these direct CP asymmetries in $D^0$ decays is
expected to be about three times smaller than calculated above.  First
measurements of relevant observables in $B^0 \to D K^{*0}$ and its
CP-conjugate have been reported very recently by the LHCb collaboration
\cite{LHCbK*0}.  Early measurements of these processes have been performed
by {\sc BaBar}~\cite{BabarK*0}.

\section*{Acknowledgements}

We thank Tim Gershon and Sheldon Stone for useful communications.
B. B. would like to acknowledge the hospitality of the Particle Theory
Group, University of Chicago.  This work was supported in part by NSERC
of Canada (BB, DL) and by the United States Department of Energy
through Grant No.\ DE FG02 90ER40560 (JLR).


\begin{thebibliography}{99}

\bibitem{CKMfitter} J. Charles {\it et al.} (CKMfitter Collaboration),
Eur.\ Phys.\ J. C {\bf 41}, 1 (2005), periodic updates at {\tt
http://ckmfitter.in2p3.fr/}.

\bibitem{UTfit} A.~J.~Bevan {\it et al.} (UTfit Collaboration),
  PoS ICHEP {\bf 2010}, 270 (2010)
  [arXiv:1010.5089 [hep-ph]].

\bibitem{GLW} M. Gronau and D. London, Phys.\ Lett.\ B {\bf 253}, 483 (1991);
M. Gronau and D. Wyler, Phys.\ Lett.\ B {\bf 265}, 172 (1991).

\bibitem{Buccella:1994nf}
  F.~Buccella, M.~Lusignoli, G.~Miele, A.~Pugliese, and P.~Santorelli,
  Phys.\ Rev.\ D {\bf 51}, 3478 (1995)
  [hep-ph/9411286];
  S.~Bianco, F.~L.~Fabbri, D.~Benson and I.~Bigi,
  Riv.\ Nuovo Cim.\  {\bf 26N7}, 1 (2003)
  [hep-ex/0309021];
  Y.~Grossman, A.~L.~Kagan and Y.~Nir,
  Phys.\ Rev.\ D {\bf 75}, 036008 (2007)
  [hep-ph/0609178].

\bibitem{ADS}
D. Atwood, I. Dunietz, and A. Soni, Phys.\ Rev.\ Lett.\ {\bf
78}, 3257 (1997); Phys.\ Rev.\ D {\bf 63}, 036005 (2001).

%
\bibitem{GGSZ} A. Giri, Y. Grossman, A. Soffer, and J. Zupan, Phys.\ Rev.\
D {\bf 68}, 054018 (2003).

\bibitem{delAmoSanchez:2010ji}
  P.~del Amo Sanchez {\it et al.} ({\sc BaBar} Collaboration),
  Phys.\ Rev.\ D {\bf 82}, 072004 (2010)
  [arXiv:1007.0504 [hep-ex]].

\bibitem{Lees:2013zd}
  J.~P.~Lees {\it et al.} ({\sc BaBar} Collaboration),
arXiv:1301.1029 [hep-ex].

\bibitem{Abe:2006hc} 
  K.~Abe {\it et al.}  (Belle Collaboration),
  Phys.\ Rev.\ D {\bf 73}, 051106 (2006)
  [hep-ex/0601032].

\bibitem{Trabelsi:2013uj} K. Trabelsi, arXiv:1301.2033, presented on behalf of
the Belle Collaboration at CKM 2012 Conference, Cincinnati, OH, 2012.

\bibitem{Aaltonen:2009hz}
  T.~Aaltonen {\it et al.} (CDF Collaboration),
  Phys.\ Rev.\ D {\bf 81}, 031105 (2010)
  [arXiv:0911.0425 [hep-ex]].

\bibitem{Aaij:2012kz}
  R.~Aaij {\it et al.} (LHCb Collaboration),
  Phys.\ Lett.\ B {\bf 712}, 203 (2012)
  [Erratum-ibid.\ B {\bf 713}, 351 (2012)]
  [arXiv:1203.3662 [hep-ex]].

 \bibitem{LHCbCONF} LHCb Collaboration, Report No.\ LHCb-CONF-2012-032,
presented at CKM 2012 Conference, Cincinnati, OH, 2012.

\bibitem{Malde:2013kf}
  S.~Malde,
  arXiv:1301.0279 [hep-ex], presented on behalf of the LHCb
Collaboration at CKM 2012 Conference, Cincinnati, OH, 2012.

\bibitem{LHCbK*0}
 R. Aaij {\it et al.} (LHCb Collaboration),
  arXiv:1212.5205 [hep-ex].

\bibitem{LHCbasym} R. Aaij {\it et al.} (LHCb Collaboration), Phys.\ Rev.\
Lett.\ {\bf 108}, 111602 (2012).

\bibitem{CDFasym} T. Aaltonen {\it et al.} (CDF Collaboration), Phys.\ Rev.\
Lett.\ {\bf 109}, 111801 (2012).

\bibitem{Wang:2012ie} 
  W.~Wang,
  Phys.\ Rev.\ Lett.\  {\bf 110}, 061802 (2013)
  [arXiv:1211.4539 [hep-ph]].

\bibitem{MZ} M. Martone and J. Zupan, arXiv:1212.0165.

\bibitem{BGR} B. Bhattacharya, M. Gronau, and J. L. Rosner, Phys.\ Rev.\ D
{\bf 85}, 054014 (2012).

\bibitem{Bhattacharya:2012kq}
  B.~Bhattacharya, M.~Gronau and J.~L.~Rosner,
 presented by M. Gronau, Proceedings of the
 Tenth International Conference on Flavor Physics and CP Violation -
 FPCP2012, May 21-- 25 2012, Hefei, SLAC eConf C120521
 [arXiv:1207.0761 [hep-ph]].

\bibitem{Bhattacharya:2012pc}
  B.~Bhattacharya, M.~Gronau and J.~L.~Rosner,
presented by B. Bhattacharya at Charm 2012, Honolulu, Hawaii, May 2012,
  arXiv:1207.6390 [hep-ph].

\bibitem{GG} M. Golden and B. Grinstein, Phys.\ Lett.\ B {\bf 222}, 501 (1989).

\bibitem{Sav} M. J. Savage, Phys.\ Lett.\ B {\bf 257}, 414 (1991).

 \bibitem{Bigi:2011em}
  I.~I.~Bigi and A.~Paul,
  J. High Energy Phys.\ 03 (2012) 021
 [ arXiv:1110.2862 [hep-ph]];
  G.~Isidori, J.~F.~Kamenik, Z.~Ligeti and G.~Perez,
  Phys.\ Lett.\ B {\bf 711}, 46 (2012)
  [arXiv:1111.4987 [hep-ph]];
 J.~Brod, A.~L.~Kagan and J.~Zupan,
  Phys.\ Rev.\ D {\bf 86}, 014023 (2012)
  [arXiv:1111.5000 [hep-ph]];
  D.~Pirtskhalava and P.~Uttayarat,
  Phys.\ Lett.\ B {\bf 712}, 81 (2012)
  [arXiv:1112.5451 [hep-ph]];
  H.~Y.~Cheng and C.~W.~Chiang,
  Phys.\ Rev.\ D {\bf 85}, 034036 (2012)
[arXiv:1201.0785 [hep-ph]];
T.~Feldmann, S.~Nandi and A.~Soni,
  J. High Energy Phys.\ 06 (2012) 007
  [arXiv:1202.3795 [hep-ph]];
H.~-n.~Li, C.~-D.~Lu and F.~-S.~Yu,
  Phys.\ Rev.\ D {\bf 86}, 036012 (2012)
  [arXiv:1203.3120 [hep-ph]];
 E.~Franco, S.~Mishima and L.~Silvestrini,
  J. High Energy Phys.\ 05 (2012) 140
  [arXiv:1203.3131 [hep-ph]];
  J.~Brod, Y.~Grossman, A.~L.~Kagan and J.~Zupan,
  J. High Energy Phys.\ 10 (2012) 161
  [arXiv:1203.6659 [hep-ph]];
H.~-Y.~Cheng and C.~-W.~Chiang,
  Phys.\ Rev.\ D {\bf 86}, 014014 (2012)
  [arXiv:1205.0580 [hep-ph]].

\bibitem{Gronau:2002mu}
  M.~Gronau,
  Phys.\ Lett.\ B {\bf 557}, 198 (2003)
  [hep-ph/0211282].

\bibitem{Belleasym} B. R. Ko, presented on behalf of the Belle Collaboration
at 36th International Conference on High Energy Physics (ICHEP2012),
Melbourne, arXiv:1212.1975.

\bibitem{HFAG} Y. Amhis {\it et al.} (Heavy Flavor Averaging Group),
arXiv:1207.1158, periodic updates at
{\tt http://www.slac.stanford/edu/xorg/hfag}.

\bibitem{CDFacp} T. Aaltonen {\it et al} (CDF Collaboration), Phys.\ Rev.\
D {\bf 85}, 012009 (2012).

\bibitem{Lees:2012jv} J. P. Lees {\it et al.} ({\sc BaBar} Collaboration),
arXiv:1212.3003.

\bibitem{Ko:2012uh}
  B.~R.~Ko {\it et al.} (Belle Collaboration),
J. High Energy Phys.\ 02 (2013) 098
  [arXiv:1212.6112 [hep-ex]].

\bibitem{Bhattacharya:2008ke}
  B.~Bhattacharya and J.~L.~Rosner,
  Phys.\ Rev.\  D {\bf 79}, 034016 (2009)
  [Erratum-ibid.\  D {\bf 81}, 099903 (2010)]
  [arXiv:0812.3167 [hep-ph]].

\bibitem{Bhattacharya:2009ps}
  B.~Bhattacharya and J.~L.~Rosner,
  Phys.\ Rev.\ D {\bf 81}, 014026 (2010)
  [arXiv:0911.2812 [hep-ph]].

  \bibitem{Isi}  I. Dunietz,
Phys.\ Lett.\ B {\bf 270}, 75 (1991).

\bibitem{Gronau:2007xg}
  M.~Gronau,
  Int.\ J.\ Mod.\ Phys.\ A {\bf 22}, 1953 (2007)
  [arXiv:0704.0076 [hep-ph]].

  \bibitem{BabarK*0}
  B.~Aubert {\it et al.} ({\sc BaBAr} Collaboration),
  Phys.\ Rev.\ D {\bf 80}, 031102 (2009)
  [arXiv:0904.2112 [hep-ex]].

\end{thebibliography}
\end{document}